# Toward Quantitative Measurements of Piezoelectricity in III-N Semiconductors Nanowires

L. Jaloustre[1,2], S. Le-Denmat[1], T. Auzelle[3], M. Azadmand[3,4], L. Geelhaar[3], F. Dahlem[2,*] and R. Songmuang[1,*]

[1]Univ. Grenoble Alpes, CNRS, Grenoble INP, Institut Néel, 38000 Grenoble, France

[2]Univ. Lyon, Ecole Centrale de Lyon, Laboratoire de Tribologie et Dynamique des Systèmes CNRS-UMR5513, 36 avenue Guy de Collongue, 69134 Écully, France

[3]Paul-Drude-Institut für Festkörperelektronik, Leibniz-Institut im Forschungsverbund Berlin e.V., Hausvogteiplatz 5-7, 10117 Berlin, Germany

[4]L-NESS and Dipartimento di Scienza dei Materiali, Università di Milano – Bicocca, Via R. Cozzi 55, 20125 Milano, Italy

*Corresponding authors: rudeesun.songmuang@neel.cnrs.fr, franck.dahlem@cermav.cnrs.fr

ABSTRACT

Piezoelectric semiconductor III-Nitride nanostructures have received increasing interest as an alternative material for energy harvesters, sensors, and self-sustainable electronics, demanding well-clarification of their piezoelectric behavior. Despite the feasibility of piezoresponse force microscopy (PFM) to resolve piezo-responses at the nanoscale, several difficulties arise when the measurements are performed on low piezo-coefficient materials due to various artifacts. This work shows that semi-quantitative PFM on low piezo-coefficient III-Nitrides can be achieved in high-aspect-ratio nanostructures such as nanowires or nanorods. For conventional bulks and thin films, accurate determination of their piezoresponses is limited because of clamping and bending effects which can occur simultaneously during PFM measurements. While the clamping effect only reduces the piezoresponse amplitude, the bending motion either increases or decreases this amplitude and can also rotate the phase by 180°. Improved electric field distribution in nanowires minimizes both artifacts, allowing correct determinations of crystal polarities and piezo-coefficients. In contrast to the reports in the literature, we do not observe giant piezoelectricity in III-N nanowires with a diameter in the range of 30-80 nm. This work provides an access to fundamental parameters for developing III-N based piezoelectric nano-devices.

INTRODUCTION

Piezoelectric properties in nanostructures that convert small mechanical displacement into electrical charges and vice-versa are considered as an alternative route toward eco-friendly energy harvesters, sensors, and self-sustainable electronics. In particular, non-centrosymmetric semiconductor nanowires (i.e. ZnO and GaN) have gained rising attention for mechanical energy scavenging applications after the report of ZnO nanowire-nanogenerators and self-powered nano-systems [1-2]. Improved electromechanical conversion efficiency was expected in ultra-small nanowires due to the enhanced piezoelectricity theoretically predicted in ZnO and GaN nanowires with a diameter smaller than 2.4 nm [3]. This giant piezoelectricity was experimentally deduced from piezoresponse force microscopy (PFM) signals of GaN single nanowires with a diameter of 60 nm, suggesting the piezoelectric coefficient up to six times the value of their bulk counterpart [4-5]. The PFM measurements on ZnO and GaN nanobelts [6-7] as well as GaN nanowires [8] also indicated their enhanced piezoelectricity. On the contrary, the piezoelectric constant of ZnO nanorods interpreted from the PFM signals was close to that of bulk values [9].

PFM is a scanning force microscopy (SFM), typically used for probing piezoelectric response in ferroelectric and piezoelectric materials at the nanoscale [10]. In conventional PFM, a conductive tip is used as a movable top electrode to apply an alternated electric field and to simultaneously detect surface displacement associated with the inverse piezoelectric effect. This displacement is usually in the picometer-range, requiring accurate detection via lock-in techniques. While the PFM amplitude is corresponding to the piezoelectric coefficients of the materials, the phase signal reveals their crystal polarities. Nevertheless, it remains challenging to quantitatively interpret the PFM signal; particularly, from piezoelectric semiconductors such as III-N and ZnO because of their low piezoelectric coefficient values (2-10 pm/V). Various parameters can introduce artifacts in the piezoresponse signals, such as electrostatic effect [11], nanoscale SFM tip diameter [12], substrate bending [13], substrate clamping [14], cantilever mechanical resonance signal amplification [15], cantilever buckling [15], topographic variation, etc. These artifacts lead to incorrect determinations of piezoelectric coefficients and material polarities. For example, those higher-than-bulk piezoelectric coefficients observed in nanowires and nanobelts [4-8] might be merely caused by the artifacts.

Here, we explore a possibility to reach *quantitative* PFM on III-N piezoelectric semiconductors in different forms, such as bulks, thin films, and nanowires, which remains controversial in the literature. Our studies reveal that the artifacts such as electrostatic, bending, and clamping effects can strongly deviate the piezoresponses (both amplitude and phase) from theoretical values. While the electrostatically induced artifacts can be minimized by using high stiffness cantilevers [11, 16] and/or top electrodes [11], the bending and the clamping effects are efficiently suppressed by reducing the lateral dimension of the studied materials because of the improved field homogeneity. Our experimental results and explanations are supported by two-dimensional (2D) finite element simulations. This work shows the potential of using high-aspect-ratio structures such as nanowires, nanorods, or nanocolumns for semi-quantitative PFM measurements that should be useful for various piezo-materials [II-VI (CdS, ZnO), III-As, PZT]. Finally, contrary to the report in ref [3,4,5], our overall analyses point out that there is no giant piezoelectric effect in III-N nanowires with a diameter of 30-80 nm.

III-Nitride wurtzite semiconductors are non-centrosymmetric crystals in a space group P63mc and a point group 6 mm. It has a six-fold rotation symmetry about the c-axis, i.e. the [0001] direction. By convention, the [0001] direction is defined by a vector pointing from a Ga atom to a nearest-neighbor N atom. This material family has both spontaneous ($P_{sp}$) and piezoelectric ($P_{pz}$) polarizations. The spontaneous one, which is the polarization at zero strain, originates from their non-ideal crystal structure [17, 18]. The orientation of this polarization is always antiparallel to the [0001] direction [Left image of Figure 1(a)] and cannot be rotated by the application of an external electric field. The piezoelectric or strain-induced polarization is caused by the center of gravity's displacement of cations and that of anions under external stress. The middle image of Figure 1(a) schematically shows the Ga-polar GaN structures with corresponding spontaneous and piezoelectric polarizations under a tensile force along the c-axis. Ionic displacements in III-N crystals occur preferably by a distortion of the tetrahedral angle rather than by a modification of the tetrahedral distance. The tensile force brings the Ga atoms more distance from the nitrogen basal atoms, causing a *positive* piezoelectric polarization which is opposite to the orientation of the spontaneous polarization. Similarly, an external electric field can also introduce ionic displacements in III-N crystals. When a positive electric field is applied along the *c*-axis of the metal-polar III-N crystals, the cations (anions) move in the same (opposite) direction to the field [19], resulting in the crystal expansion along this direction [Right image of Figure 1(a)]. In the case of a negative applied field, the behavior is the opposite.

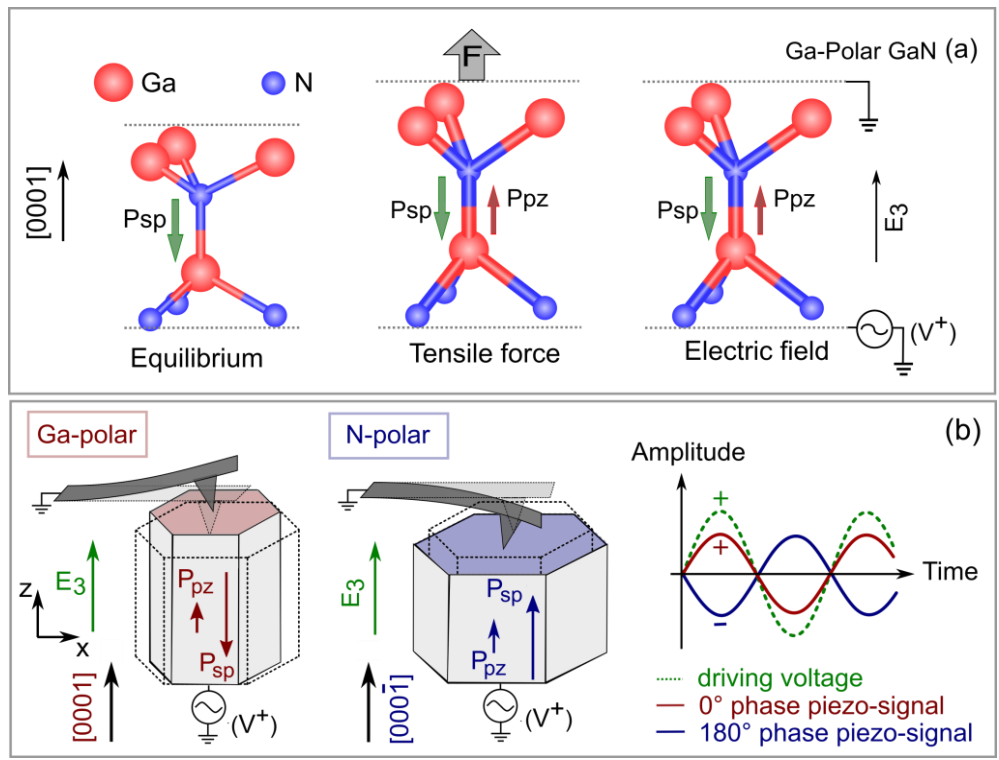

FIGURE 1 Schematic illustrations of (a) wurtzite Ga-polar GaN crystal structures with corresponding spontaneous ($P_{sp}$) and piezoelectric ($P_{pz}$) polarizations at equilibrium (Left), under a tensile force along the [0001] direction (Middle), and under a positive applied field along the [0001] direction (Right). (b) Expected piezoresponses from Ga-and N-polar GaN bulk samples with an alternating voltage applied to the bottom surface. The presented images are the positive cycle of sinusoidal bias. The coordinate of the system is defined such that three orthogonal axes of the measurements (the axes x, y, and z, respectively) are aligned with the [11-20], [-1100], and [0001] directions of Ga-polar GaN, while the z-axis is along the [000-1] direction of N-Polar GaN.

Figure 1(b) schematically illustrates the expected piezoresponses from Ga- and N-polar GaN bulk samples with an alternating voltage applied to the bottom surface. The coordinate of the system is defined such that three orthogonal axes of the measurements (the axes $x$, $y$, and $z$, respectively) are aligned with the [11-20], [-1100], and [0001] directions of Ga-polar GaN, while the $z$-axis is along the [000-1] direction of N-Polar GaN. The $x$-and the $y$-axes are rotated following the right-hand system. The positive cycle of the sinusoidal bias in this configuration introduces a positive electric field ($\boldsymbol{E}_z = E_3 \cdot \boldsymbol{z}$) in the [0001] direction of Ga-polar GaN, while this field is antiparallel to the [0001] direction of N-polar one. For an ideally uniform $E_3$, the strain in the [0001] direction ($\varepsilon_{33}$) of III-N is given by $\varepsilon_{33} = d_{33} \cdot E_3$, where $d_{33}$ is the piezoelectric coefficient defined with respect to this crystallographic axis. In the case of metal-polar III-N crystals; i.e. $d_{33} > 0$ [20], they expand ($\varepsilon_{33} > 0$) under a positive $E_3$, and contract ($\varepsilon_{33} < 0$) when the electric field is negative. The crystal behavior is reversed for N-polar III-N which has $d_{33} < 0$. These scenarios are presented in the left and middle panels of Figure 1(b), respectively. The relation between the strain tensor and the components of the displacement vectors allows us to extract piezoelectric constants from the surface displacement detected by PFM [21]. Because of the crystal symmetry of GaN, the displacement along the c-axis ($u_3$) is given by $u_3 = d_{33} \cdot E_3 \cdot h = d_{33} \cdot V$, where $h$ is the sample thickness and $V$ is the applied voltage across the substrate. The expected phase difference between the vertical PFM signal and the alternating voltage applied to the bottom surface [the right panel of Figure 1 (b)] is 0° for Ga-polar GaN (the red curve), while it shifts to 180° for N-polar GaN (the blue curve). In principle, this PFM phase response can be used as a non-destructive method to determine crystal polarities.

In 2002, Rodriguez *et al.* reported a polarity imaging of a one-µm thick GaN layer grown on a sapphire substrate by monitoring the PFM phase signal [22, 23]. In that article, it was stated that the piezoelectric polarization in GaN is parallel to the spontaneous one for the tensile strain [22,23]. This statement is correct only when the crystal is biaxially tensiled in the plane perpendicular to the [0001] direction [24], which causes a compressive strain in the [0001] direction due to Poisson's effect. Incorrect description of the strain-induced polarization caused a mistake in phase interpretation. The reported PFM signal in ref [22,23] actually has a *180° phase shift* from the piezo-displacement expected from GaN. Here, we emphasize that the piezoelectric and spontaneous polarization are parallel when the GaN is compressed (contract) in the [0001] direction, while both polarizations are antiparallel if the crystal is tensiled (expand) in this direction [see Figure 1]. Stoica *et al.* [25] described the field-induced crystal displacement by using the spontaneous polarization, showing that their observed PFM phase signal also has a 180° phase rotation. In 2011, Brubaker *et al.* noticed that the PFM phase signal from a 100-nm AlN layer on Si(111) agrees with the piezo-induced displacement [26, 27], unlike the ones from GaN reported in ref [22, 23, 25]. In 2015, Minj *et al.* used the PFM phase to locate Ga- and N-polar regions in single nanowires [28], following the questionable phase interpretations of ref [22-23]. For ZnO which has a similar wurtzite crystal structure to GaN, Scrymgeour *et al.* showed that the PFM phase signal of ZnO bulk corresponds to the piezo-induced response [29]. In contrast, the PFM phase of ZnO thin films measured by Guillemin *et al.* [30], was implicitly coherent with the spontaneous polarization direction. None of those authors, except Brubaker *et al.*, has pointed out the discrepancies of those PFM phase interpretations. Thus, systematic and thorough PFM investigations are necessary for quantitative evaluations of piezoelectric coefficients and material polarities of III-N piezoelectric semiconductors. The in-depth understanding of PFM measurements could also clarify the origin of the giant piezoelectricity in nanowires and nanobelts found by this technique.

EXPERIMENTAL SECTION

For our PFM investigations, the samples were 400-µm thick Ga- and N-polar GaN single crystal bulk substrates from LumiLog. 500-nm thick Ga-and N-polar GaN films on 200 nm-AlN buffer layers grown on highly doped Si(111) substrates, as well as a 200-nm Al-polar AlN layer grown on a highly doped Si(111) substrate from Kyma technology, were also measured. On each sample, 5-10 nm Ti/20nm Au circular top electrodes with a diameter ranging from 250 to 1000 µm were deposited. The metal/semiconductor contact has Schottky characteristics that might affect the PFM signal at high voltage bias. These situations must be further investigated. A 250-nm thick non-piezoelectric $SiO_2$ layer on a highly doped Si substrate and a commercial PZT bulk PCI300 from PI Ceramics with a known polarization direction were used as reference samples.

The nanowire samples were grown by using plasma-assisted molecular beam epitaxy (PAMBE). Undoped GaN nanowires and undoped $Al_{0.9}Ga_{0.1}N$/GaN nanowires were deposited on highly n-doped Si(111) substrates [31, 32], while undoped AlN nanowires were grown on a sapphire substrate covered by a conductive TiN layer obtained by sputtering [33]. These nanowires are single crystal and nearly defect-free. The carrier concentration of the undoped GaN nanowires was in the range of $10^{17}$ cm$^{-3}$ [34]. The polarity of the GaN nanowires is of N-polar, proven by KOH chemical etching [35], and deduced from high-resolution scanning transmission electron microscopy (HR-STEM) of the equivalent nanowires [36]. The AlN nanowires were suggested to have Al-polarity by HR-STEM analyses of one dispersed AlN nanowire from the same sample [33]. Details of the investigated nanowires are summarized in Table I.

Table 1: Details of undoped nanowires for PFM studies in this work.

|  | GaN |  | AlGaN/GaN | AlN |
|---|---|---|---|---|
| Length (nm) | 550 | 1000 | 180/600 | 800 |
| Diameter (nm) | 30-50 | 30-50 | 60 | 80 |
| Polarity | N-Polar [35,36] |  |  | Al-Polar [33] |
| Carrier concentration (cm$^{-3}$) | $10^{17}$ |  | Insulating/ $10^{17}$ | Insulating |

In the case of a top electrode on an ensemble of undoped GaN nanowires of a 1-µm length, the nanowires were firstly embedded inside Polymethyl Methacrylate (PMMA). Then, $O_2$ plasma etching was applied to remove the PMMA layer that covered the top of the nanowires. Afterward, a circular 10-nm Ti/ 60-nm Pt top electrode with a diameter of 1-mm was deposited. All samples were glued on metallic holders by using silver epoxy (EPOTEK: H21D) or silver paste which also served as a back electrode.

The PFM measurements were performed by using Bruker D3100-SFM, with external AC and DC excitations and a lock-in detection via Zurich Instrument HF2LI. The resolution limit of our setup is ~ 0.5 pm determined by PFM background noises measured on non-piezoelectric substrates such as Silicon. The PtIr, Ti/Ir, or Cr/Pt coated SFM conductive cantilevers have spring constants of 0.2, 3, and 40 N/m (PPP-EFM, ASYELEC.01-R2, or ElectriAll-In-One), with typical contact resonant frequencies of around 70, 350, and 1400 kHz, respectively. Note that this frequency varied from one cantilever to the others, thus it should be measured before each PFM measurement. The frequency dependence of the PFM signal was measured to ensure that there is no mechanical amplification at the selected measurement frequency (17 kHz in our case). Our PFM amplitude and phase were generally constant in the frequency range of 5-50 kHz. The holders of the cantilever-tips and the samples have homemade electrical connectors available for electrical excitations. The alternated voltage bias with the amplitude value of $V_m$ was applied to the substrate backside. The SFM tips and the top electrodes were connected to the ground via external metallic wires using an external connection box and a micro-bonding technique, respectively. The external wiring for electrical connection is more reliable than using only a tip-electrode mechanical connection which can cause unstable PFM signals [37]. A non-local electrostatic contribution was determined via a contact potential difference ($V_{CPD}$) obtained by an open-loop Kelvin Probe Force Microscopy (KPFM) in a standard intermittent mode [38].

The piezoresponse amplitudes were extracted from the X-signal instead of the R-signal of the lock-in amplifier. We also verified that the Y-signal was nearly zero, indicating a minimized background contribution [39]. The PFM phase was determined by the *sign* of the X-value. The positive and negative signals present the piezo-induced surface displacement which is in-phase and 180° out-of-phase with the excitation voltage, respectively. The experimental piezoelectric constants ($d_{33}^{exp}$) were extracted from the slope of the X-signal versus the excitation amplitude, $V_m$, while the signs of these slopes; ideally, should provide the sample polarity. This $d_{33}^{exp}$ is an apparent piezoelectric constant extracted from the measured surface displacement which is the intrinsic piezo-induced displacement modified by the effects of the measurement configurations and the artifacts.

RESULTS AND DISCUSSION

Figure 2(a) summarizes the vertical PFM X-signal as a function of $V_m$ from N-polar GaN bulk samples without any top electrodes. The measurements were performed by using cantilevers with different spring constants, i.e., 0.2 N/m, 3 N/m, and 40 N/m. The vertical X-signals measured by using the 0.2 and 3 N/m cantilevers are in-phase with the excitation voltage and have a slope ($d_{33}^{exp}$) of about +8 pm/V and +1 pm/V; respectively. This in-phase signal is opposite to the expected phase of the piezoresponse from N-polar GaN. The inconsistency is ascribed to a dominant electrostatic contribution in the detected PFM signal. The electrostatic force often occurs between the SFM cantilever-tip and the sample surface because of surface potential and surface charges [40]. This force has the same frequency as the excitation voltage; thus, it cannot be filtered by lock-in techniques. It can either attract or repulse the cantilever, causing the movements that super-impose with the ones induced by the crystal displacement. In our particular case, the electrostatic force counterbalances the cantilever motion. We emphasize that if the sign of $d_{33}^{exp}$ which should reveal the material polarity is not correct, the absolute value of $d_{33}^{exp}$ cannot be directly used for determining the piezo-coefficients. The PFM artifacts can indeed determine this value.

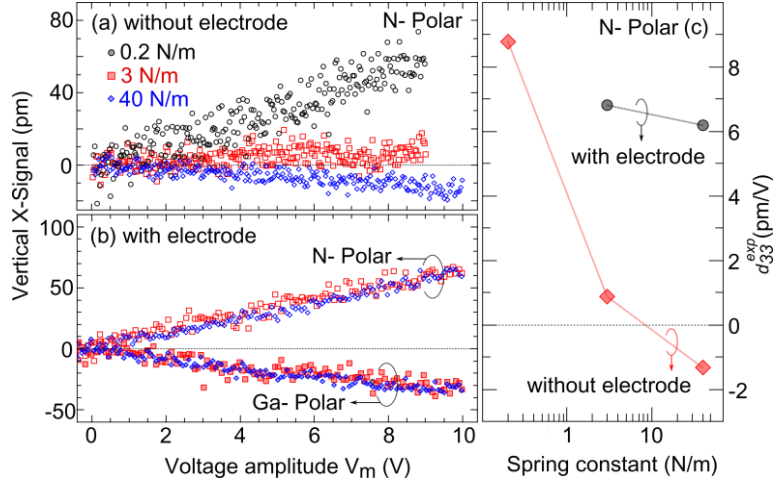

FIGURE 2 *Vertical X-signal as a function of voltage excitation amplitude ($V_m$) taken from (a) an N-polar GaN bulk sample without any top electrodes and (b) the center of 500-µm diameter top electrodes of N- and Ga-polar GaN bulk samples. These values were measured by using the cantilever stiffnesses of 0.2 N/m (○), 3 N/m (□), and 40 N/m (◊). (c) $d_{33}^{exp}$ from N-polar GaN bulk samples with and without top electrodes plotted as a function of spring constants.*

When using 40 N/m cantilevers which are less sensitive to electrostatics, the $d_{33}^{exp}$ value is -1.2 pm/V, which means that the PFM signal has a phase shift of 180° from the excitation voltage. This phase signal agrees with the expected piezo-response of N-polar GaN, implying that the electrostatic effect is less dominant. The absolute value of $d_{33}^{exp}$ is lower than the theoretical one of GaN bulk substrate ($d_{33}^{theo}$=2.29 pm/V) [20], attributed to the clamping effect. Such a clamping is caused by the electrically-unexcited surrounding material which limits the mechanical deformation of the excited volume due to the highly focused field underneath the SFM tip [41]. In the case of thin films on Si, the inactive Si substrate also participates in the clamping. Moreover, this highly tip-focused field might also localize in a non-piezoelectric oxide that forms on the surface of the piezoelectric material. As a consequence, the piezo-material is exposed to a lower electric field than the applied one. Both effects can significantly decrease the PFM amplitude but cannot rotate the phase by 180°.

Besides using the stiff cantilevers, it is possible to reduce the electrostatic effect by applying a direct voltage $V_{DC} = V_{CPD}$ during the PFM measurements [16]. However, this method is not straightforward for semiconductors because the $V_{CPD}$ depends on the surface band bending, which varies with the excitation voltage. The application of a fixed $V_{DC}$ might not be sufficient to cancel the electrostatic potential at different excitation voltage. An alternative to suppress the electrostatic effect is to deposit the metallic electrode on the sample surface, and electrically connect it to the cantilever [11]. When the tip is in contact with the top electrode, this configuration equalizes the potential of the cantilever-tip with that of the sample surface, consequently suppressing the electrostatic effect. Figure 2(b) shows the vertical X-signal as a function of $V_m$ taken at the center of 500-µm diameter top electrodes on Ga- and N-polar GaN bulk samples measured by 3 and 40 N/m cantilevers. A priori, the electrode center of these samples is a symmetric-point, where the in-plane displacement does not contribute to the vertical motion of the cantilever [42]. Thus, the vertical signal directly corresponds to the vertical displacement of the crystal surface without any artifacts due to the cantilever buckling motion. In this figure, there is no significant difference between the PFM responses obtained by using these two cantilever stiffnesses, reflecting the electrostatic shielding of the top electrode. Nevertheless, the phase signals of both samples were rotated by 180°, relative to the piezo-induced one of the respective bulk materials. For example, we obtained the phase signal of 0° (instead of 180°) from the N-polar GaN bulk sample, while the phase signal from the Ga-polar one was 180° (instead of 0°). This opposite phase indicates that the observed PFM signals relate to the material polarity. Nevertheless, they cannot be directly described by the piezo-displacements as previously discussed in the ideal case shown in Figure 1. This 180° phase offset is not because of the electrostatic force since it was entirely eliminated by the use of the top electrode and the high stiffness cantilever. Figure 2(c) summarizes $d_{33}^{exp}$ from the N-polar GaN bulk samples without and with the top electrodes plotted as a function of cantilever stiffnesses. The figure reveals that the correct sign of $d_{33}^{exp}$; which is negative for N-polar GaN, is obtained only when the 40 N/m cantilever was used on the samples without any top electrodes.

Table 2: Summary of the absolute value and the sign of $d_{33}^{exp}$ extracted from the PFM measurements of III-N bulk samples and thin films with different electrode diameters, in comparison with $d_{33}^{theo}$. The grey shades highlight the cases in which the sign of $d_{33}^{exp}$ agrees with $d_{33}^{theo}$.

| Samples | Polarity | Electrode diameter | $d_{33}^{exp}$ from PFM | | $d_{33}^{theo}$ | |
|---|---|---|---|---|---|---|
| | | | $\lvert d_{33}^{exp} \rvert$ (pm/V) | Sign | $\lvert d_{33}^{theo} \rvert$ (pm/V) | Sign* |
| Thin films | Ga-Polar | SFM tip | 1.7 | Positive | 2.29 [20] | Positive |
| | | 250 µm | 2.39 | Negative | | |
| | | 500 µm | 2.8 | Negative | | |
| | N-Polar | SFM tip | 0 | N.C. | 2.29 [20] | Negative |
| | | 250 µm | 0.7 | Positive | | |
| | | 500 µm | 1.09 | Positive | | |
| | | 1000 µm | 2.52 | Positive | | |
| | Al-Polar | SFM tip | 1.24 | Positive | 5.35 [20] | Positive |
| | | 250 µm | 0.3 | Positive | | |
| | | 500 µm | 0.45 | Negative | | |
| Bulk samples | Ga-Polar | SFM tip | 0.2 | Positive | 2.29 [20] | Positive |

|  |  | 250 µm | 3.1 | Negative |  |  |
|---|---|---|---|---|---|---|
|  |  | 500 µm | 3.5 | Negative |  |  |
|  | N-Polar | SFM tip | 0.45 | Negative | 2.29 [20] | Negative |
|  |  | 250 µm | 4.6 | Positive |  |  |
|  |  | 500 µm | 6.5 | Positive |  |  |

*The sign of $d_{33}^{theo}$ corresponds to the condition in which a positive electric field is applied in the [0001] direction of wurtzite III-N.

The $d_{33}^{exp}$ taken at the electrode center from GaN bulk samples and thin films on Si(111) are summarized and plotted as a function of electrode diameters in Table 2 and Figure 3, respectively. Systematically, the absolute value of $d_{33}^{exp}$ decreases when the electrode diameter is reduced, while their sign changes to *the opposite* when the SFM tip was used as a top electrode. This sign inversion could be a signature of the bending motion of the substrate, which can reverse the piezoresponse by 180° once it is dominant [13]. Despite the sign which is coherent with the material polarity when the tip is used as the top electrode (as highlighted in grey in Table 2), the $d_{33}^{exp}$ is lower than the theoretical bulk values ($d_{33}^{theo}$), which are the intrinsic value of unclamped materials. This observation is explained by the highly focused field only beneath the tip induces a clamping effect which limits the mechanical deformation of the excited volume.

We have also performed the PFM measurements on the samples with 10-100 µm electrode diameters. However, these electrodes cannot be directly connected to the ground by external wiring due to the limitation of the bonding setup. The electrodes were mechanically contacted to the metallic SFM tip which was grounded through its holder. Ideally, when the metal-coated cantilever is in mechanical contact with the metal electrode, they should have the same electrical potential that suppresses the electrostatic contribution. However, the metallic surfaces are usually covered by the oxide layers which inhibit a stable electrical connection between the tip and electrodes [37], leading to unreliable measurements of $d_{33}^{exp}$.

This $d_{33}^{exp}$ variation cannot be attributed to the electrostatic contribution as it was already suppressed or minimized. In the case of the metallic electrode, the electrostatic contribution was nullified by electrically connecting both cantilever-tip and electrode to the ground. When the cantilever-tips were used for electrical excitation, we selected the 40N/m cantilevers which are much less sensitive to the electrostatic force [see Figure 2(a)]. Besides, this observation cannot be described by the electrode thickness variation because its thickness is much smaller than that of piezo-material. When the thicknesses of the electrode and the piezo-material are comparable, the in-plane deformation of the piezo-material could introduce the in-plane one of the top electrodes, resulting in the change of the electrode thickness due to Poisson's effect [43]. Then, the $d_{33}^{exp}$ value deduced from the vertical displacement could be affected.

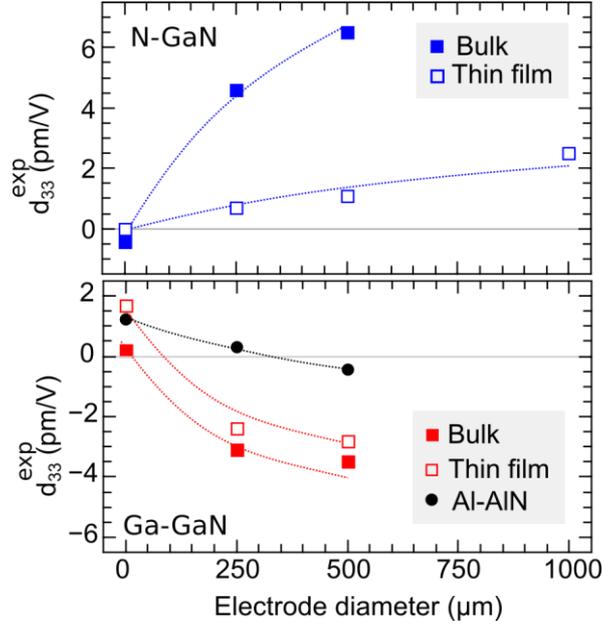

FIGURE 3 Summary of $d_{33}^{exp}$ measured at the electrode center plotted as a function of top electrode diameters. We assume that the excitation diameter of the SFM tip is 1 nm. These PFM measurements were performed on Ga- and N-polar GaN bulk samples and thin films on Si(111), as well as a 200 nm Al-polar AlN layer on Si(111). The dashed lines are the guide for the eyes.

To qualitatively support our interpretations of the PFM signals, two-dimensional (2-D) finite element simulations were performed by using TiberCAD and COMSOL, taking advantage of six-fold rotational symmetry. The electric field-induced displacement was calculated by solving the constitutive equations of piezoelectricity in a linear regime. The simulated structure was a Ga-polar GaN bulk substrate with a 400-µm thickness and a 6000-µm width, equivalent to those of our studied samples. The diameter of a top electrode was varied from 200 nm to 6000 µm, while the substrate backside was entirely covered by a metallic electrode. The electrode thickness, the effects of free carriers, and the bandstructures were not taken into account in our simulations. For the electrical boundary conditions, a static voltage of 1 V was applied to the substrate backside whereas the potential of the top electrode was set to zero. For the mechanical boundary conditions, two simulated scenarios were applied, i.e., the fully clamped and the unclamped substrate backsides. The substrate sidewalls were set as unclamp. The properties of GaN and AlN used for modeling are summarized in Table 3.

We consider that the static simulation is sufficient to describe our experimental results. The PFM phase and amplitude are not frequency-dependent in the excitation range of 5-50 kHz, indicating that the excitation frequency of 17 kHz is far from the internal mechanical mode of the samples. The static simulations provide a snapshot of the dynamic motion of the crystal surface, i.e., the surface displacement derived from a positive (negative) static voltage could represent the surface crystal movement in a positive (negative) cycle of the alternated voltage bias. Note that the static bending possibly caused by the static force exerted by the probe is not taken into account here, because it was filtered out by the lock-in amplifier that detects only the displacement signal of 17 kHz.

TABLE 3. Metal-polar III-N Material properties used for finite element simulations.

| Material properties | GaN | AlN | References |
|---|---|---|---|
| Elastic constants | $C_{11}$=370 GPa<br>$C_{12}$=145 GPa<br>$C_{13}$=110 GPa<br>$C_{33}$=390 GPa<br>$C_{44}$= 90 GPa | $C_{11}$=396 GPa<br>$C_{12}$=137 GPa<br>$C_{13}$=108 GPa<br>$C_{33}$=373 GPa<br>$C_{44}$=116 GPa | [20] |
| Piezoelectric coefficients | $e_{15}$= -0.3 C/m$^2$<br>$e_{31}$ -0.34 C/m$^2$<br>$e_{33}$= 0.67 C/m$^2$ | $e_{15}$= -0.48 C/m$^2$<br>$e_{31}$ -0.58 C/m$^2$<br>$e_{33}$=1.55 C/m$^2$ | [20] |
| Dielectric constants | $\varepsilon_{11}$ = 9.5<br>$\varepsilon_{33}$ =10.4 | $\varepsilon_{11}$ = 9<br>$\varepsilon_{33}$ =10.7 | [44] |

Figures 4(a)-(b) are the contour plots showing the vertical displacement in the [0001] direction of a Ga-polar GaN bulk substrate with a 500-µm diameter top electrode, simulated by setting unclamped and clamped backside, respectively. The deformation is plotted with a scaling factor of 5x10$^7$, to be visible in the figures. For the unclamped substrate, the vertical displacement of the crystal surface below the electrode moves to negative which is opposite to the one of the clamped substrate. The sectional profiles of the vertical displacement taken from the unclamped Ga-polar GaN bulk with a top electrode diameter of 10-µm and 500-µm are plotted in comparison in Figures 4(c)-(d). The red solid-lines are the top surface displacement while the black solid-lines represent the bottom surface one. The piezoelectric constants were calculated by dividing the vertical displacement of the top surface taken at the electrode center by the applied voltage. These values from the clamped ($d_{33}^{clamp}$) and unclamped ($d_{33}^{unclamp}$) substrates are plotted versus the top electrode diameters in Figure 4(e).

By increasing the electrode diameter, the electric field becomes better homogeneous across the substrate thickness. For the clamped substrate [the blue circle symbols in Figure 4(e)], the $d_{33}^{clamp}$ converges to the $d_{33}^{theo}$ of 2.29 pm/V when the electrode diameter is beyond ~1000 µm in this particular geometry. Below this critical diameter, the $d_{33}^{clamp}$ decreases at a smaller electrode diameter because the displacement of the excited volume is limited by the unexcited surrounding material due to a localized field. Note that the sign of $d_{33}^{clamp}$ is always positive, as expected for Ga-polar GaN. As the substrate backside is mechanically fixed, i.e. the displacement at the bottom surface is set to zero, the top surface displacement is always positive (expands) under a positive field.

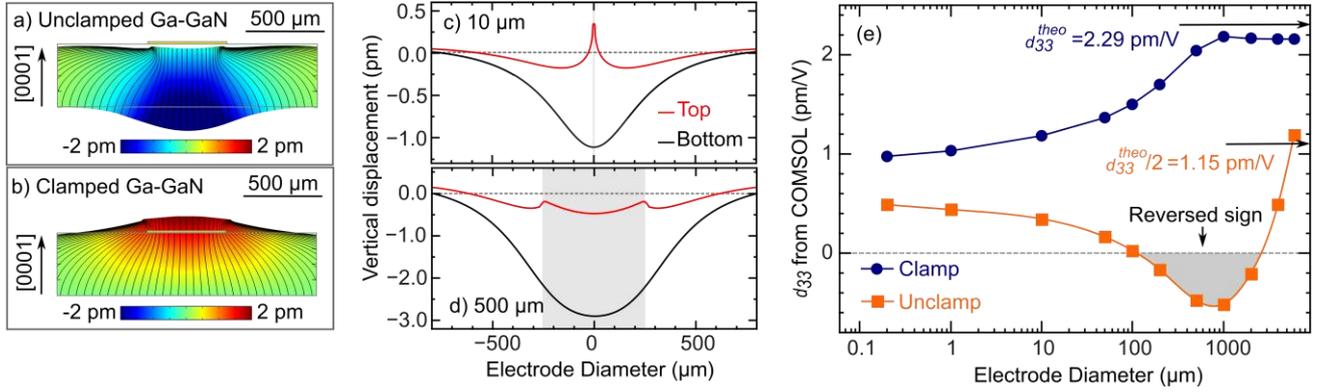

*FIGURE 4 (a)-(b) Contour plots from the simulations showing the vertical displacement in the [0001] direction of a Ga-polar GaN bulk substrate with a top electrode diameter of 500 µm. The substrate backside was biased by a static voltage of 1 V, while the top electrode was set at 0 V. The black solid lines present the line field distribution. The substrate deformation is shown with a scaling factor of 5x10$^7$. The positions of the top electrode are indicated by the yellow thin-rectangles at the center of the top surface. (c)-(d) Sectional profiles of the vertical displacement of the top (the red line) and the bottom surface (the black line) of an unclamped Ga-polar GaN bulk substrate with a top electrode diameter of 10 µm and 500 µm, respectively. The grey shade defines the area inside the top electrode. (e) $d_{33}^{clamp}$ (the blue circle symbols) and $d_{33}^{unclamp}$ (the orange square symbols) plotted as a function of electrode diameters.*

Unlike the clamped boundary condition, the $d_{33}^{unclamp}$ evolution is also influenced by the bottom-surface displacement besides the clamping effect [see Figure 4(a), (c), and (d)]. We separate this curve into three parts. The first one where the top electrode diameter is larger than ~ 3000 µm, the $d_{33}^{unclamp}$ increases as a function of electrode diameters. This is because the applied field becomes homogeneously distributed across the substrate, consequently reducing the influences of the bottom-surface displacement and the clamping effect on the $d_{33}^{unclamp}$. Once their dimension of the top electrode and the substrate are equal at 6000 µm, the top and bottom crystal surfaces would symmetrically displace in opposite directions. The $d_{33}^{unclamp}$ analyzed only from the top surface displacement is thus close to $d_{33}^{theo}/2$. In the second part where the electrode diameter is in the range of 100 - 2500 µm [the grey area in Figure 4(e)], the $d_{33}^{unclamp}$ is negative. In this region, the bottom surface displacement plays an important role, resulting in the dominant substrate bending. When the electrode diameter is less than 100 µm, the field is more localized at the top surface, the $d_{33}^{unclamp}$ returns a positive value as the bottom surface motion is less significant. However, it remains lower than $d_{33}^{theo}/2$ due to the restricted vertical displacement by a larger volume of unexcited material.

These simulations and our experiments of position-dependent vector PFM [42] support the interpretation that the bending motion is responsible for the 180° phase shift (the sign inversion) of the detected vertical displacement signal shown in Figure 3. From the vector PFM measurements [42], the lateral and the buckling motions of the cantilever detected at different positions inside the electrode, reveal that the lateral displacement of the crystal surface corresponds to the bending motion of the substrates. Those measurements were performed on both N-Polar and Ga-Polar GaN crystals. The experimental details are fully described in ref [42]. It is possible that gluing the samples to the holder by using silver epoxy or silver paste in our experimental conditions, may not perfectly clamp the substrate backside. The bowing of the substrates might also be one of the reasons for the imperfect clamping of the samples. When the electrode diameter is below a critical value, the bending displacement is minimized but the clamping effect becomes superior. We emphasize that the situation depends on various parameters such as sample (substrate and contact) geometries, mechanical characteristics of the samples and gluing materials, location of the tip in respect to the sample boundaries, and sample holders [45, 46, 47, 48, 49]. The bending effect was found by other measurement techniques that access only the upper surface displacement, e.g. single interferometer [50] and laser doppler vibrometer [51, 52].

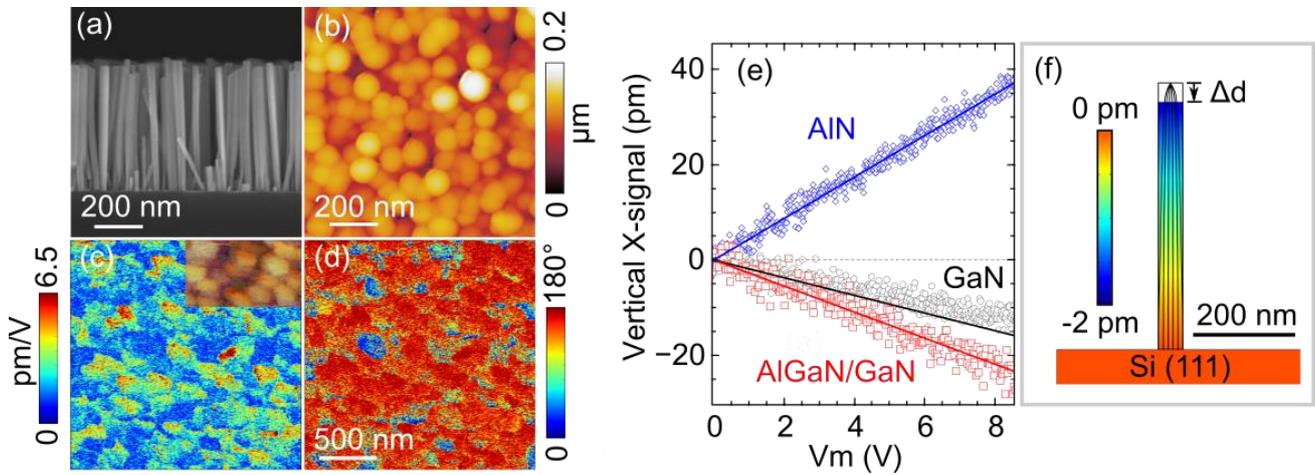

FIGURE 5 (a) Side-view SEM image and (b) top-view SFM topographic image measured by using an intermittent mode of a typical N-polar GaN nanowire ensemble. (c)-(d) 2x2 µm² scanning images of the $\left|d_{33}^{exp,NW}\right|$ (extracted from the vertical X-amplitude) and the phase of N-polar GaN nanowires excited by $V_m$=6 V at 9 kHz. The inset shows the corresponding SFM topography simultaneously acquired during the PFM measurements. (e) Vertical X-signal from GaN (○), AlN (◊), and $Al_{0.9}Ga_{0.1}N/GaN$ (□) nanowires plotted as a function of the voltage amplitude $V_m$. The solid lines present the vertical displacement versus the applied voltage of each material from the simulations. (f) Contour plot of vertical displacement of an N-polar GaN nanowire on Si substrate. The top electrode with a 1-nm diameter is set at 0 V while the voltage of 1 V is applied at the Si substrate surface. The black solid lines represent the field distribution inside the nanowire.

It was suggested that reducing the width-to-height ratio of the studied materials can release the constraint imposed by inactive-surrounding materials, thus minimizing the clamping effect [43, 53, 54]. Here, we performed the PFM measurements on III-N nanowires such as GaN, $Al_{0.9}Ga_{0.1}N/GaN$, and AlN, with an aspect ratio of about 1:10. Figures 5(a)-(b) are the side view SEM image and the top view SFM topographic image measured by using an intermittent (tapping) mode of a typical undoped GaN nanowire ensemble. The PFM measurements were performed on the top of these nanowires.

Figures 5(c)-(d) are 2x2 µm$^2$ scanning images of the vertical X-amplitude and the phase of the undoped N-polar GaN nanowires obtained by setting V$_m$= 6 V at 9 kHz on a highly doped Si substrate. The 3 N/m cantilevers were selected for these measurements as they are sufficiently flexible for SFM contact-scanning on the top of the nanowires, providing less topological artefacts [the inset of Figure 5(c)]. This image was simultaneously obtained during the scanning PFM measurement. The PFM signals are highly visible in the nanowire regions and generally have a *180° out-of-phase* with the electrical excitation. This result agrees with the expected phase of the piezo-induced displacement of N-polar GaN and other characterizations of equivalent nanowires such as KOH etching [35] and HR-STEM [36]. A few nanowires show a phase signal of 0°, which might reflect their Ga-polarity. This observation is consistent with the fact that the wires grown on Si(111) by PAMBE appear to be Ga-polar at 90% [35]. The higher-than-bulk PFM amplitude in this particular image is attributed to a higher background at this selected excitation frequency. When the PFM scanning image is taken only at a fixed excitation voltage, the piezo-coefficient deduced from this PFM signal is not accurate as it might include the background level. For this work, the PFM scanning measurements were performed mainly to provide the distribution of the PFM signal in the nanowire ensemble.

To be more quantitative, the vertical PFM X-signal as a function of the applied voltage V$_m$ were measured on the top surface of the nanowires of GaN, Al$_{0.9}$Ga$_{0.1}$N/GaN, and AlN as shown in Figure 5(e). Several measurements were performed at different positions on the sample surface, using the 40 N/m cantilevers which are less electrostatic sensitive. These high stiffness cantilevers are not suitable for contact scanning SFM on the flexible structures as the resulting images are usually distorted. The phase responses of the N-polar GaN nanowires and the N-polar Al$_{0.9}$Ga$_{0.1}$N/GaN nanowires are always *180° out-of-phase*, confirming the scanning phase image in Figure 5(d). On the other hand, the positive PFM signal of the AlN nanowires corresponds to the *in-phase* response, agreeing with the metal polarity observed by HR-STEM of one dispersed AlN nanowire from the same sample [33]. The absolute value of piezocoefficients from the N-polar GaN nanowires ($|d_{33}^{exp,NW}|$) were in the range of 1-1.6 pm/V, while the higher values of 2.4-3.6 pm/V, were systematically found in the Al$_{0.9}$Ga$_{0.1}$N/GaN nanowires. For the AlN nanowires, the $|d_{33}^{exp,NW}|$ were in the range of 4.4-4.5 pm/V. The larger $d_{33}^{exp,NW}$ fluctuation of the Al$_{0.9}$Ga$_{0.1}$N /GaN nanowires is attributed to the alloy variation in the Al$_{0.9}$Ga$_{0.1}$N sections [32]. Unlike the results of the N-polar bulk samples without top electrodes shown in Figure 2(b), the sign inversion of the $d_{33}^{exp,NW}$ was not found when changing the cantilever stiffness from 3 N/m to 40 N/m. Both cantilever stiffnesses provide a correct phase signal of the nanowires, indicating that the piezo-induced displacement is more important than the electrostatically-induced one since the clamping effect is released in such a geometry. To entirely screen the electrostatic effect, the 1-µm long N-polar GaN nanowire ensemble was covered by a one-mm diameter metallic electrode. Their PFM phase response indicates their N-polarity with the $|d_{33}^{exp,NW}|$ of 1.5-2 pm/V, slightly higher than those of the nanowires without top electrode. Table 4 presents the summary of $d_{33}^{exp,NW}$ extracted from PFM measurements (the grey columns) which are generally lower than the values of $d_{33}^{theo}$ of respective bulk materials. In contrast to the reports in the literature [4, 5], we do not observe higher-than-bulk piezo-coefficients or giant piezoelectricity in PAMBE grown III-N nanowires with a diameter in the range of 30-80 nm. The table also shows the agreement between the polarities interpreted from the sign of $d_{33}^{exp,NW}$ and the ones deduced from other characterization techniques, in all samples.

Table 4: Comparison between $d_{33}^{exp,NW}$ extracted from PFM measurements with $d_{33}^{NW}$ from the simulations and $d_{33}^{theo}$. The corresponding structural parameters for simulations are summarized in the table. The polarities of the nanowire interpreted from the sign of $d_{33}^{exp,NW}$ are compared with the ones deduced from other characterization techniques.

| Nanowires | PFM | | Polarity (Other characterizations) | Theoretical bulk value $\|d_{33}^{theo}\|$ (pm/V) | Nanowire simulations | | |
|---|---|---|---|---|---|---|---|
| | $\|d_{33}^{exp,NW}\|$ (pm/V) | Polarity (the sign of $d_{33}^{exp,NW}$) | | | Diameter, Length (nm) | Electrode diameter* (nm) | $\|d_{33}^{NW}\|$ (pm/V) |
| Undoped GaN | 1-1.6 | N (negative) | N[a] | 2.29[c] | 50, 550 | 1 | 1.87 |
| | | | | | | 50 | 2.14 |
| Undoped GaN with a top electrode | 1.5-2 | N (negative) | N[a] | 2.29[c] | 50, 1000 | 1 | 1.96 |
| | | | | | | 50 | 2.15 |
| | | | | | | 50** | 1.88** |
| Al$_{0.9}$Ga$_{0.1}$N/ GaN | 2.4-3.6 | N (negative) | N[a] | 3.0[d] | 50, 780 | 50 | 2.65[d] |
| AlN | 4.4-4.5 | Al (positive) | Al[b] | 5.35[c] | 80, 1000 | 1 | 4.36 |
| | | | | | | 80 | 4.94 |

[a] Experimental results from ref [35,36].

[b] Experimental results from ref [33]

[c] Data from ref [20].

[d] The values are obtained by using equations (1) and (2).

*The 1-nm diameter top electrode mimics the contact area between the SFM tip diameter and the nanowire. This electrode diameter was increased to the nanowire diameter to imitate the situation where the nanowire top surface is fully covered by a metal electrode.

**The nanowire is encapsulated in PMMA.

We performed finite element simulations of single GaN and Al$_{0.9}$Ga$_{0.1}$N/GaN nanowires with an N-polarity and an AlN nanowire with an Al-polarity on a Si(111) substrate, applying a fully-clamped boundary condition at the backside of a Si substrate. The Si substrate was treated as an isotropic material with Young's modulus of 185 GPa and Poisson's ratio of 0.14 [55]. The nanowire dimensions for the simulations were selected to be close to the studied samples as summarized in Table 4. The diameter and the length of an undoped GaN nanowire were 50 nm and 550 nm, respectively. The ones of the GaN nanowire encapsulated in PMMA were 50 nm and 1000 nm. For the $Al_{0.9}Ga_{0.1}N/GaN$ nanowire, the diameter was set at 50 nm, while the length of the $Al_{0.9}Ga_{0.1}N$ section was 180 nm and that of the GaN stem was 600 nm. The simulated structure of the AlN nanowires was 80 nm in diameter and 800 nm in length. The top electrode diameter was set at 1 nm, to mimic the contact area between the SFM tip diameter and the nanowire. This electrode diameter was increased to the nanowire diameter to imitate the situation where the nanowire top surface is fully covered by a metal electrode. For electrical boundary conditions, the top electrode was set to 0 V while the top and bottom surfaces of the Si substrate is set to 1 V to represent a highly doped substrate (highly conductive).

The piezoelectric coefficients of $Al_xGa_{1-x}N$ on GaN $[d_{33}(Al_xGa_{1-x}N\ on\ GaN)]$ of bulks and nanowires were calculated by using the following linear interpolations.

$$d_{33}(Al_xGa_{1-x}N\ on\ GaN) = d_{33}(Al_xGa_{1-x}N) \cdot \frac{L_{Al_xGa_{1-x}N}}{L_{tot}} + d_{33}(GaN) \cdot \frac{L_{GaN}}{L_{tot}} \quad (1)$$

$$d_{33}(Al_xGa_{1-x}N) = (1-x) \cdot d_{33}(GaN) + x \cdot d_{33}(AlN) \quad (2)$$

, where $x$ is the Al content in $Al_xGa_{1-x}N$ alloy. $L_{GaN}$ and $L_{Al_xGa_{1-x}N}$ are the thicknesses of GaN and $Al_xGa_{1-x}N$, respectively, while $L_{tot} = L_{GaN} + L_{Al_xGa_{1-x}N}$ is the total thickness of both layers. $d_{33}(Al_xGa_{1-x}N)$, $d_{33}(GaN)$, and $d_{33}(GaN)$ are the piezoelectric coefficients of $Al_xGa_{1-x}N$, GaN, and AlN in the forms of bulks and nanowires. For simplicity, we considered that $Al_xGa_{1-x}N$ on GaN is fully relaxed.

The solid lines in Figure 5(e) present the simulated vertical displacements at the center of the nanowire top surface as a function of the applied voltage $V_m$, in comparison with the PFM results. The piezoelectric constants of the nanowires from the simulations ($d_{33}^{NW}$) are the slopes of these curves. These values are summarized in Table 4 with their corresponding structural parameters. It is evidenced that $d_{33}^{NW}$ and $d_{33}^{exp,NW}$ are all consistent, approaching $d_{33}^{theo}$. In comparison to bulk samples and thin films, the bending and clamping effects are reduced in the nanowire geometry ascribed to the improved field distribution in the material. Nevertheless, the higher absolute values of the $d_{33}^{NW}$ when the electrode diameter was increased from 1 nm to the nanowire diameter (either 50 nm or 80 nm), suggests that there remains a clamping effect when the SFM cantilever-tips are used for electrical excitation. These results might explain a slightly lower $d_{33}^{exp,NW}$ than $d_{33}^{theo}$ of the clamp-free bulk materials. The simulations also show that the PMMA encapsulating layer could constrain the piezo-displacement of the nanowires, slightly reducing the $d_{33}^{NW}$ despite their much lower Young's modulus of around 3 GPa. Nevertheless, the variation of piezo-coefficients in this range is difficult to experimentally detect because of the resolution limit of the PFM setup.

CONCLUSIONS

This work shows how PFM measurement artifacts (electrostatics, bending, and clamping) can significantly mislead the interpretation of the PFM signal, in particular, for the materials with low piezoelectric constants such as III-N materials. The electrostatic contribution on the PFM signals of GaN bulk samples and thin films can be minimized either by using high stiffness cantilevers and/or by using top metallic electrodes. Depending on the electrode and sample geometries, the bending and clamping effects can influence the PFM amplitude and phase signal, giving incorrect piezocoefficient values and polarities. Our results and simulations reveal that both effects are diminished in high-aspect-ratio structures such as nanowires. Hence, the promising way to reach quantitative PFM measurements is to reduce the lateral dimension of the measured materials and to use the stiff cantilevers. In our work which follows such a protocol, the polarities of III-N nanowires extracted from the PFM signals are consistent with the expectations and other measurement techniques, while the $d_{33}^{exp,NW}$ values are in agreement with the $d_{33}^{NW}$ from finite element simulations, which are close to the theoretical bulk values, $d_{33}^{theo}$. From our extensive PFM measurements of III-N bulk samples, thin-film, and nanowires, we suggest that the giant piezoelectric effect does not exist in the GaN, AlN, and AlGaN/GaN nanowires equivalent to those studied in ref [4,5]. We also propose that the interpreted Ga-polar areas in GaN nanowires shown by Minj *et al.* in ref [28] might be N-polar ones. The present studies should provide a way to access piezoelectric properties of nano-materials which are fundamental parameters for the new generation of energy harvesters, sensors, and self-sustainable electronics.

ACKNOWLEDGMENTS

The authors acknowledge the assistance from Nanofab/Néel Institute for sample fabrications. The work is financially supported by the ARC-Energy project, the NanoTribelec-ANR project, GaNEX, and the German Bundesministerium für Bildung und Forschung (Project No. FKZ:13N13662).